\documentclass[12pt,english,floatfix,nofootinbib,superscriptaddress,aps,prd,preprint]{revtex4}
\usepackage[utf8]{inputenc}
\usepackage{float}
\usepackage{array}
\usepackage{lipsum}
\usepackage{graphicx}
\usepackage{amsmath,amsthm,amsfonts,amssymb}
\usepackage{graphicx}
\usepackage[english]{babel}
\usepackage{color}
\usepackage{tensor}
\usepackage{esint}
\usepackage[dvips]{epsfig}
\usepackage[dvips]{graphicx}
\usepackage{float}
\usepackage{units}
\usepackage{textcomp}
\usepackage{mathrsfs}
\usepackage{amsmath}
\usepackage[makeroom]{cancel}
\usepackage{amssymb}
\usepackage{amsbsy}
\usepackage{amsfonts}
\usepackage{amssymb,mathrsfs,xcolor}
\usepackage{esint}
\usepackage{array}
\usepackage{graphicx}

\usepackage{hyperref}
\hypersetup{
    colorlinks,
    citecolor=blue,
    filecolor=green,
    linkcolor=purple,
    urlcolor=red,
}

\usepackage{slashed}

\newcommand{\ie}{\begin{equation}}
\newcommand{\fe}{\end{equation}}
\newcommand{\se}{\begin{eqnarray}}
\newcommand{\ff}{\end{eqnarray}}

\usepackage{hyperref}
\hypersetup{colorlinks,breaklinks,
			citecolor=[rgb]{0,0.0,1.0},
            urlcolor=[rgb]{0.0,0.0,1.0},
            linkcolor=[rgb]{0,0.5,0.9}}

\begin{document}

\title{Particles in Loop Quantum Gravity formalism: a thermodynamical description}

\author{A. A. Ara\'{u}jo Filho}
\email{dilto@fisica.ufc.br}

\affiliation{Universidade Federal do Cear\'a (UFC), Departamento de F\'isica,\\ Campus do Pici,
Fortaleza - CE, C.P. 6030, 60455-760 - Brazil.}




\date{\today}

\begin{abstract}
In this work, we analyze the thermodynamical behavior of massive and massless particles within Loop Quantum Gravity formalism. We investigate a modified dispersion relation which suffices to derive all our results of interest in an \textit{analytical} manner. Initially, we study the massive case where, in essence, we examine how the mass modifies the thermodynamical properties of the system. On the other hand, to the massless case, the thermodynamical system turns out to depend only on a specific thermal state quantity, the temperature. More so, the analysis of the spectral radiation, the equation of states, the mean energy, the entropy, and the heat capacity are provided as well for both cases. Finally, the thermodynamical properties of our particles under consideration depends on the Riemann zeta function $\xi(s)$.

\end{abstract}

\maketitle


\section{Introduction}

The insight that the entropy of a black hole should be proportional to its horizon area has notably contributed to guide quantum gravity research \cite{rovelli2004quantum,kiefer2007quantum,bianchi2011polyhedra,carlip2001quantum,amelino2004severe}. Particularly, in agreement with Bekenstein \cite{bekenstein2020black}, the contribution to black hole entropy may be acquired by taking into account mainly basic elements. Moreover, the correlation involving entropy and event horizon area established notable constraint in the investigation of quantum gravity formalism.

Many attempts to reproduce the linearity of the entropy area result, utilizing straightforward quantum properties of black holes failed at the initial stages. Nevertheless, in the last decades, string theory and loop quantum gravity gave rise to some remarkable techniques that facilitated the progress in evaluating entropy when quantum properties of black holes are considered \cite{ghosh2014statistics,mansuroglu2021fermion,xiao2022logarithmic,rovelli1996black,ashtekar1998quantum,kaul2000logarithmic,strominger1996microscopic,solodukhin1998entropy,solodukhin2020logarithmic,meissner2004black}. The essential results are not restricted to the surface area contribution; they go beyond. Primarily, it is determined that the corrections in the leading order must be a log-type one. In addition, it is assumed that the relation between the entropy of a black hole and its respective event horizon takes the form:
\ie
S = \frac{A}{4 l_{P}} + \rho \ln [A/l_{P}^{2}] + \mathcal{O}(l^{2}_{P}/A).
\fe

On the other hand, within the context of Loop Quantum Gravity formalism, there does not exist a consensus on the logarithmic corrections controlled by $\rho$. Despite this, it is proposed that the correction terms, superior than the logarithimic dependence on the area, are absent \cite{ashtekar1998quantum,rovelli1996black}. In this sense, the comprehension of entropy, i.e., up to the leading log correction, could be employed to ensure the Planck scale modifications of the components within the Bekenstein study.
It seems to offer a chance to set the viability of working on some scenarios emerging from Loop Quantum Gravity, e.g., as it is shown in Ref. \cite{amelino2004severe}, where one particular modified dispersion relation is proposed-- involving a term with a linear dependence on the Planck length $l_{P}$.

Undoubtedly, a prominent aspect when one deals with a modified dispersion relation is certainly its respective thermodynamical properties ascribed to a given theory. Investigations involving thermal aspects in the context of Lorentz violation could supply further knowledge about primordial stages of expansion of the Universe. In other words, this corroborates the fact that the size at these stages are consistent with the characteristic scales of Lorentz violation \cite{kostelecky2011data}.
The thermal properties within the context of Lorentz symmetry breaking has been initially proposed by Ref. \cite{colladay2004statistical}. After that, recently many works have been made in various distinct scenarios, such as, graviton \cite{aa2021lorentz}, Pospelov and Myers-Pospelov \cite{araujo2021thermodynamic,anacleto2018lorentz} electrodynamics, CPT-even and CPT-odd violations \cite{casana2008lorentz,casana2009finite,araujo2021higher}, higher-dimensional operators \cite{reis2021thermal}, bouncing universe \cite{petrov2021bouncing,petrov2021bouncing2}, and Einstein-eather theory \cite{aaa2021thermodynamics}.

Although traditionally many indefinite studies begin with the investigation based on the action associated with a particular theory, in the literature, there exists a prominent alternative manner of addressing physical results starting exclusively from its respective dispersion relation instead \cite{amelino2001testable}. Also, in this context, there are a lack of studies examining the thermodynamic properties in the context of Loop Quantum Gravity up to now. In this sense, we follow the methodology developed in Ref. \cite{colladay2004statistical} in order to accomplish a thermodynamic description of the system. We take into account two kinds of particles: the massive and the massless ones. For all of them, we perform the analysis of the following respective state thermodynamic physical quantities: the equation of states, the mean energy, the spectral radiance, the entropy, and the heat capacity.

\section{The massive case}

The possibility of Planck-scale modifications of the dispersion relation has been extensively considered in the quantum gravity literature \cite{amelino1998tests,garay1998spacetime,amelino2002doubly,magueijo2003generalized,kowalski2003non,amelino2002doubly2,cai2014testing,myers2004experimental,arzano2016gravity,sudarsky2003bounds,calcagni2019gravitational,chen2014effects,hossenfelder2013minimal,mercati2010probing,laanemets2022observables,gong2022gravitational} and, specifically, in Loop Quantum Gravity scenario \cite{amelino12004quantum,alfaro2000quantum,smolin2002quantum,gambini1999nonstandard,ronco2016uv,bojowald2005loop,brahma2017linking,ashtekar2021short}. Although customarily many examinations begin theirs investigations with the action associated with a particular theory, there exists a prominent alternative manner in the literature of addressing physical results from its dispersion relation only \cite{amelino2001testable}. Here, such a procedure is invoked in order to provide the development of the thermodynamic properties.
This section is devoted to study massive particles within the context of Loop Quantum Gravity formalism proposed by Ref. \cite{amelino2004severe}. In this one, based on the black hole area-entropy law, it is proposed several constraints concerning Loop Quantum Gravity energy-momentum dispersion relation. Likewise, we start with the following dispersion relation for the sake of developing a procedure to acquire the thermodynamic state functions:
\ie
E \simeq {\bf{k}} + \frac{m^{2}}{2{\bf{k}}} + \alpha l_{P} E^{2},
\fe
where, $E$ is the energy, ${\bf{k}}$ is the momentum, $m$ is the mass, $\alpha$ is an arbitrary constant, and $l_{P} = \sqrt{8 \pi G}$.
Notice that the relation between energy and momentum has a different form from the usual one, as we could naturally expect. In other words, this aspect indicates that the Lorentz symmetry is no longer preserved \cite{maluf2019antisymmetric,schreck2022lorentz} in such a context. Moreover, from the above expression, we can clearly obtain two solutions; nevertheless, accomplishing a Wick-like rotation in the mass term, only one of them is in agreement with our purpose, i.e., of having a real positive definite values, which is
\ie
{\bf{k}}= \frac{1}{4} \left(2 E+\sqrt{\left(2 E-2 E^2 l_P\right){}^2+8 m^2}-2 E^2 l_P\right),
\fe
where we have considered $\alpha = 1$, and, naturally, we can derive its infinitesimal quantity $\mathrm{d} {\bf{k}}$ as follows:
\ie
\mathrm{d} {\bf{k}} = \frac{1}{4} \left(\frac{\left(2-4 E l_P\right) \left(2 E-2 E^2 l_P\right)}{\sqrt{\left(2 E-2 E^2 l_P\right){}^2+8 m^2}}-4 E l_P+2\right) \mathrm{d}E.
\label{vol}
\fe
With these above expressions, we are able to perform the integration over the momenta space in order to acquire the accessible states of the system
\ie
\Omega(E) = \frac{\Gamma}{\pi^{2}} \int^{\infty}_{0} \mathrm{d} {\bf{k}} |{\bf{k}}|^{2}, \label{ms2}
\fe
where $\Gamma$ is regarded as the volume of the thermal reservoir. Thereby, Eq. (\ref{ms2}) reads
\ie
\begin{split}
\Omega(E) = &\frac{\Gamma}{\pi^{2}} \int^{\infty}_{0}  \frac{1}{64} \left(\frac{\left(2-4 E l_P\right) \left(2 E-2 E^2 l_P\right)}{\sqrt{\left(2 E-2 E^2 l_P\right){}^2+8 m^2}}-4 E l_P+2\right) \\
& \times \left(2 E+\sqrt{\left(2 E-2 E^2 l_P\right){}^2+8 m^2}-2 E^2 l_P \right)^2 \,\mathrm{d}E.
\end{split}
\fe
It is worthy to be mentioned that all calculation present in this manuscript will be accomplished in a ``per volume'' approach. Seeking a better comprehension to the reader, we provide the most general definition of the partition function considering an indistinguishable spinless gas \cite{greiner2012thermodynamics}: 
\ie
Z(T,\Gamma,N) = \frac{1}{N!h^{3N}} \int \mathrm{d}q^{3N}\mathrm{d}p^{3N} e^{-\beta H(q,p)}  \equiv \int \mathrm{d}E \,\Omega(E) e^{-\beta E}, \label{partti1}
\fe
where, here, we have $h$ being the Planck's constant, $\beta = 1/k_{B}T$, $k_{B}$ being the Boltzmann constant, $H$ being the Hamiltonian of the system, $p$ being the generalized momenta, $q$ being the generalized coordinates, and $N$ being the number of particles. However, we notice that Eq. (\ref{partti1}) does not suffice to categorize the spin of the respective particles under consideration. For our study, such a feature has to be implemented in the following manner \cite{reis2020does,oliveira2019thermodynamic,oliveira2020relativistic,oliveira2020thermodynamic,reis2021fermions} 
\ie
\mathrm{ln}[Z] = \int \mathrm{d}E \,\Omega(E) \mathrm{ln} [ 1- e^{-\beta E}],
\fe
where the factor $\mathrm{ln} [ 1- e^{-\beta E}]$ accounts for the Bose-Einstein statistics. Thereby, we are able to derive the partition function in a straightforward way:
\ie
\begin{split}
\mathrm{ln}[Z(l_{P},m,\beta)] = -&\frac{\Gamma}{\pi^{2}} \int^{\infty}_{0}  \frac{1}{64} \left(\frac{\left(2-4 E l_P\right) \left(2 E-2 E^2 l_P\right)}{\sqrt{\left(2 E-2 E^2 l_P\right){}^2+8 m^2}}-4 E l_P+2 \right) \\
& \times \left(2 E+\sqrt{\left(2 E-2 E^2 l_P\right){}^2+8 m^2}-2 E^2 l_P \right)^2 \mathrm{ln}(1-e^{-\beta E}) \,\mathrm{d}E.
\end{split}
\fe
In possession with above expression, we are capable of inferring all thermal quantities of interest that we shall investigate in the following sections. As we shall see, in their totally, the expressions are much complicated to be solved in an \textit{analytical} way, i.e., they have \textit{numerical} solutions only. However, under a certain limit, they can properly be provided in an exact form. The definitions of the thermodynamic functions are given by
\ie
\begin{split}
 & F(\beta, m)=-\frac{1}{\beta} \mathrm{ln}\left[Z(\beta, m)\right], \\
 & U(\beta, m)=-\frac{\partial}{\partial\beta} \mathrm{ln}\left[Z(\beta, m)\right], \\
 & S(\beta, m)=k_B\beta^2\frac{\partial}{\partial\beta}F(\beta, m), \\
 & C_V(\beta, m)=-k_B\beta^2\frac{\partial}{\partial\beta}U(\beta, m),
\label{properties}
\end{split}
\fe  
where $F(\beta, m)$ is the Helmholtz free energy, $U(\beta, m)$ is the mean energy, $S(\beta, m)$ is the entropy, and $C_{V}(\beta, m)$ is the heat capacity. Furthermore,
at the begging, we initiate with the examination of the equation of states of the system, which is present in the next subsection.

\subsection{Equation of states}

This subsection is devoted to explore the consequences of the equation of states of a system composed exclusively by massive particles; to step toward to the calculation, we write:
\ie
\mathrm{d}F = - S \,\mathrm{d}T - p\,\mathrm{d}V,
\fe
which, straightforwardly, it gives rise to

\begin{figure}[tbh]
  \centering
  \includegraphics[width=8cm,height=5cm]{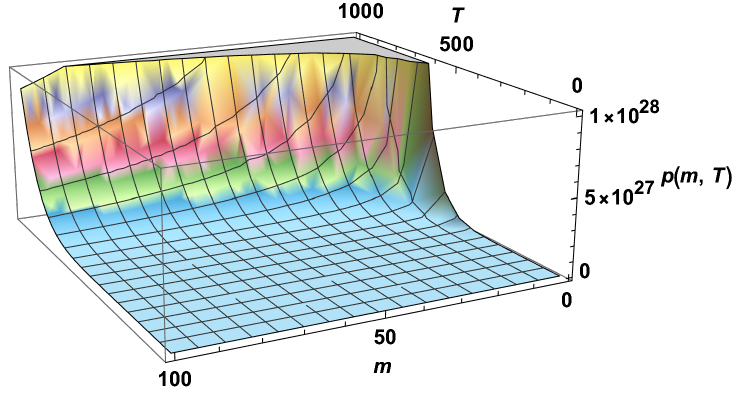}
  \includegraphics[width=8cm,height=5cm]{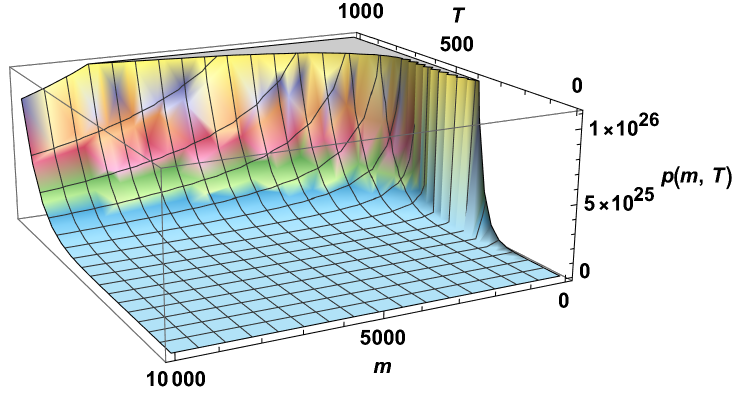}
  \includegraphics[width=8cm,height=5cm]{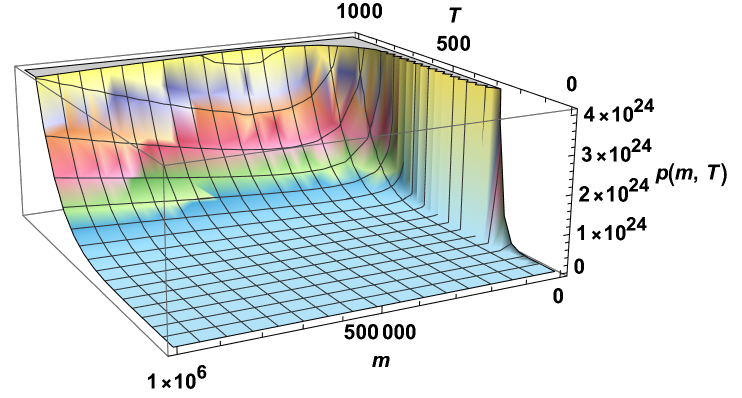}
  \includegraphics[width=8cm,height=5cm]{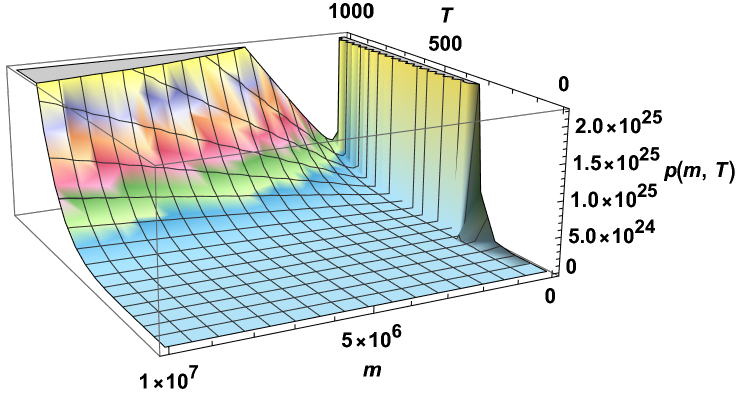}
  \caption{The equation of states for different values of mass keeping the same magnitude of temperature.}\label{equationofstatesfordifferentmass1}
\end{figure}

\begin{figure}[tbh]
  \centering
  \includegraphics[width=8cm,height=5cm]{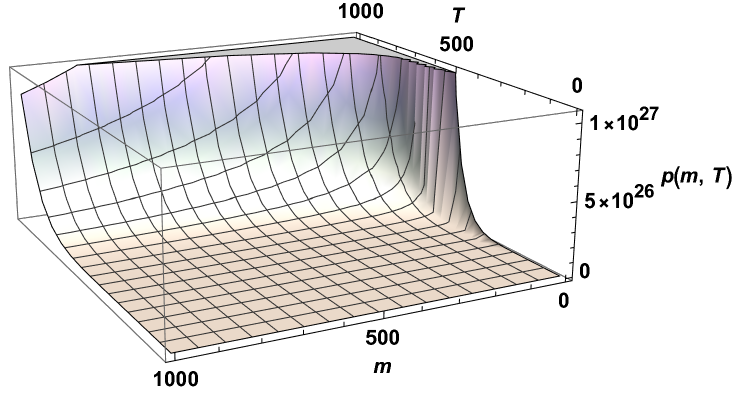}
  \includegraphics[width=8cm,height=5cm]{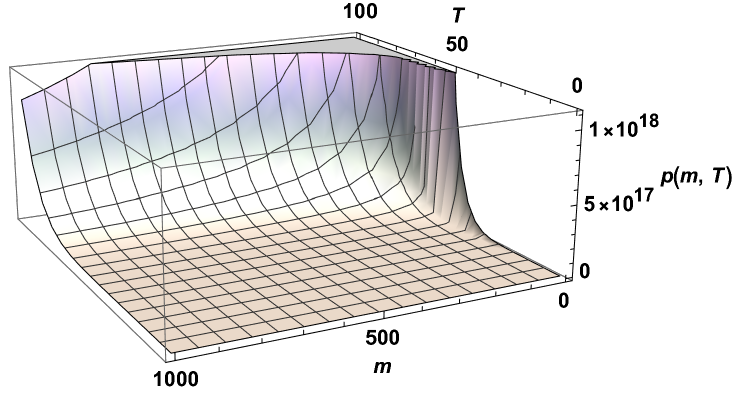}
  \includegraphics[width=8cm,height=5cm]{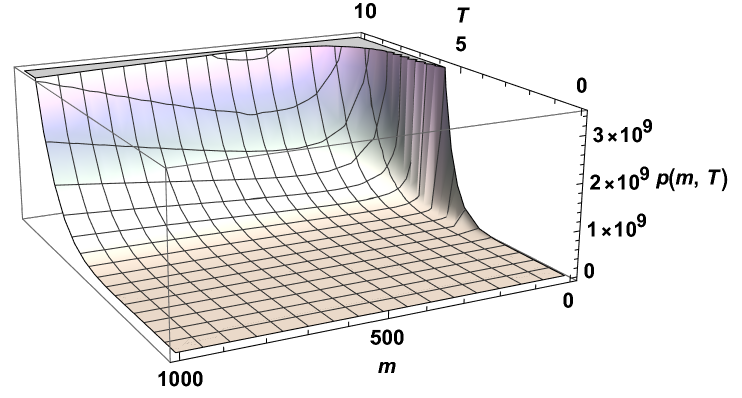}
  \includegraphics[width=8cm,height=5cm]{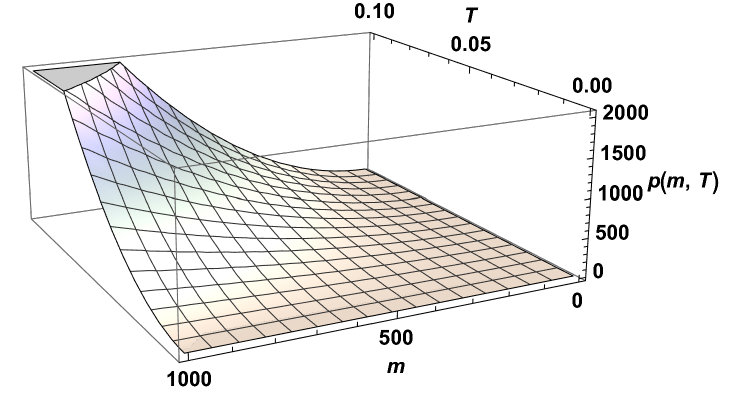}
  \caption{The equation of states for different values of temperature keeping the same magnitude of mass.}\label{equationofstatesfordifferentmass}
\end{figure}

\ie
\begin{split}
p(l_{P},m,\beta) = - \left(\frac{\partial F}{\partial V} \right)_{T} = &\frac{1}{\beta \pi^{2}} \int^{\infty}_{0}   \frac{1}{64} \left(\frac{\left(2-4 E l_P\right) \left(2 E-2 E^2 l_P\right)}{\sqrt{\left(2 E-2 E^2 l_P\right){}^2+8 m^2}}-4 E l_P+2 \right) \\
& \times \left(2 E+\sqrt{\left(2 E-2 E^2 l_P\right){}^2+8 m^2}-2 E^2 l_P\right)^2 \mathrm{ln}(1-e^{-\beta E}) \,\mathrm{d}E. \label{pressure1}
\end{split}
\fe
Since there is no exact solution to the above expression, we shall consider the limit where $\left(2 E-2 E^2 l_P\right){}^2 \ll 1$. The usage of this limit is totally reasonable since parameters which control the Lorentz symmetry breaking are  expect to be very small \cite{kostelecky2011data}.
With it, we obtain an \textit{analytical} solution to Eq. (\ref{pressure1}) as follows
\ie
\begin{split}
p(l_{P},m,\beta) = & \frac{1}{2520 \sqrt{m^{2}}\beta^{9}}\left[ 60 \beta ^2 l_P^2 \left(2 \pi ^6 \beta  \sqrt{m^2}+189 \sqrt{2} \left(\beta ^2 m^2 \zeta (5)+15 \zeta (7)\right)\right) \right.\\
& \left. - \beta ^3 l_P \left(63 \pi ^4 \sqrt{2} \beta ^2 m^2+1260 \beta  \sqrt{m^2} \left(\beta ^2 m^2 \zeta (3)+18 \zeta (5)\right)+20 \pi ^6 \sqrt{2}\right) \right.\\
& \left.-84 \beta  l_P^3 \left(2700 \beta  \sqrt{m^2} \zeta (7)+\pi ^8 \sqrt{2}\right)+1587600 \sqrt{2} \zeta (9) l_P^4 \right.\\
& \left.+21 \beta ^4 \left(5 \pi ^2 \beta ^3 \left(m^2\right)^{3/2}+\pi ^4 \beta  \sqrt{m^2} +45 \sqrt{2} \left(\beta ^2 m^2 \zeta (3)+\zeta (5)\right)\right)\right],
\end{split}
\fe
where $\zeta(s)$ is the Riemann zeta function given by
\ie
\zeta(s) = \sum^{\infty}_{n=1} \frac{1}{n^{s}} = \frac{1}{\Gamma(s)}\int_{0}^{\infty} \frac{x^{s-1}}{e^{x}-1} \mathrm{d}x, \,\,\,\,\,\,\, \Gamma(s)= \int_{0}^{\infty} x^{s-1} e^{-x} \mathrm{d}x.
\fe
In order to provide a concise analysis to this thermodynamic property, we supply its respective plot. The initial consideration presented here is that the temperature is fixed while there exists a change in the mass of the system; such behavior is shown in Fig. \ref{equationofstatesfordifferentmass1}. Notice that the system is sensitive to the change of mass, i.e., being more accentuated for higher values of $m$. On the other hand, if we regard fixed values of $m$ while the temperature varies, we will acquire the behavior exhibited in Fig. \ref{equationofstatesfordifferentmass}. In this case, up to very low temperatures though, there is no much change in the shape of the plot ascribed to the modification of the temperature.

\subsection{The mean energy}

\begin{figure}[tbh]
  \centering
  \includegraphics[width=8cm,height=5cm]{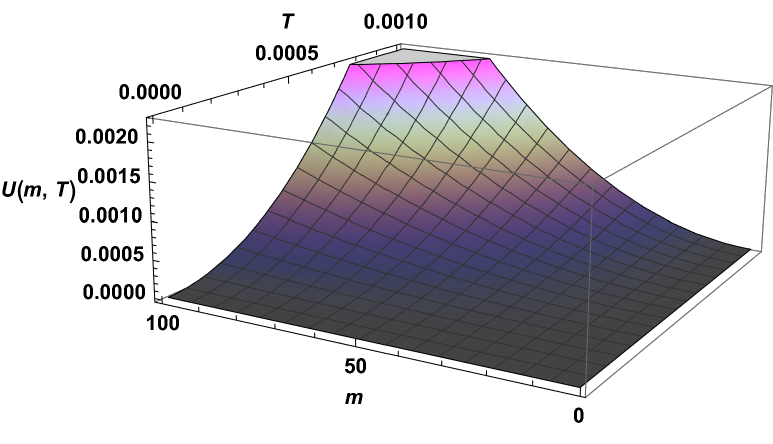}
  \includegraphics[width=8cm,height=5cm]{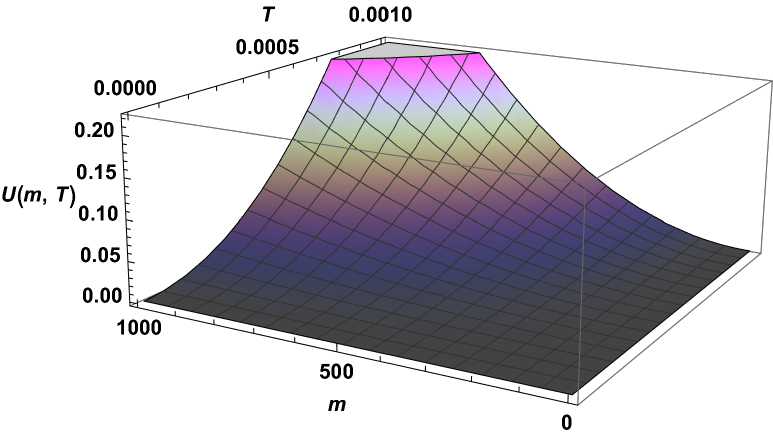}
  \includegraphics[width=8cm,height=5cm]{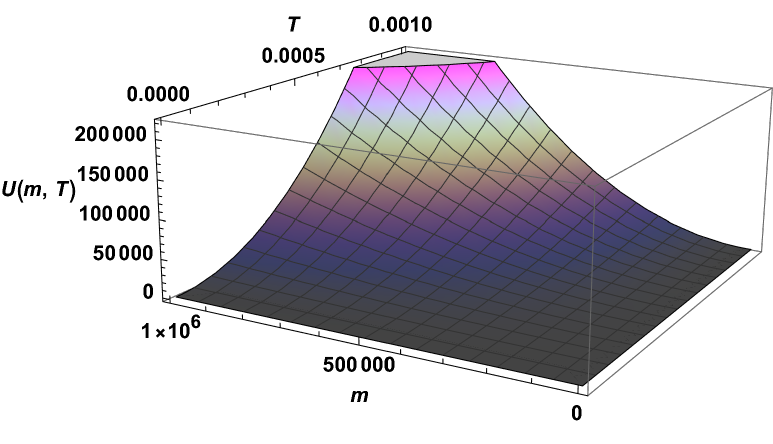}
  \includegraphics[width=8cm,height=5cm]{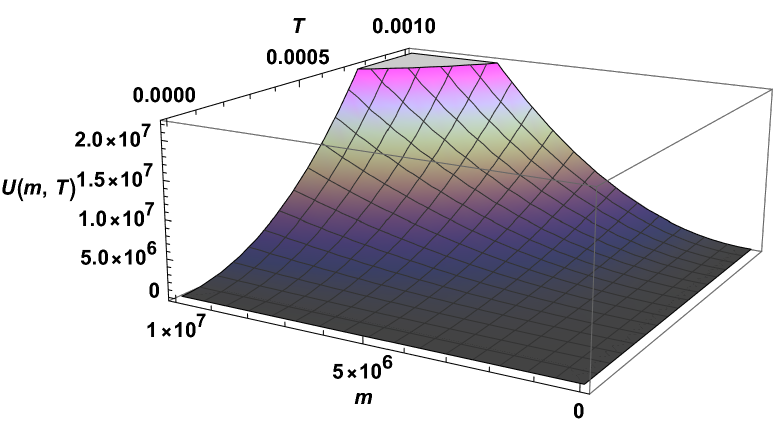}
  \caption{The mean energy for different values of mass maintaining the same magnitude of temperature}\label{ufordifferentmass}
\end{figure}

\begin{figure}[tbh]
  \centering
  \includegraphics[width=8cm,height=5cm]{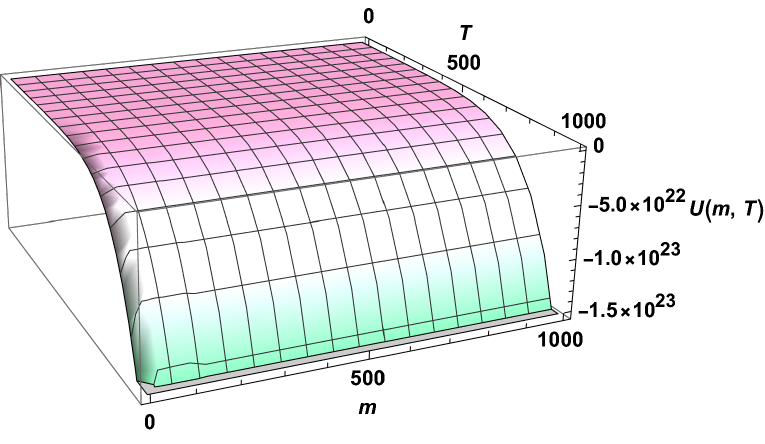}
  \includegraphics[width=8cm,height=5cm]{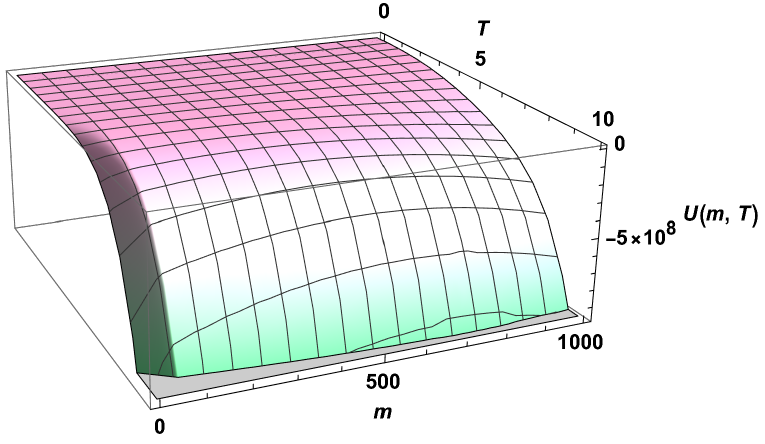}
  \includegraphics[width=8cm,height=5cm]{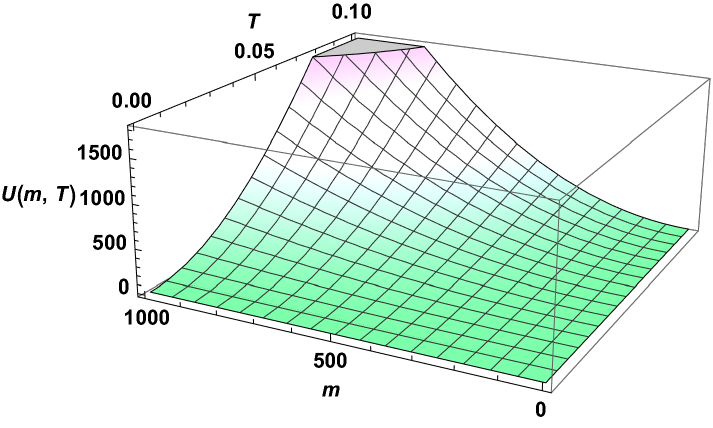}
  \includegraphics[width=8cm,height=5cm]{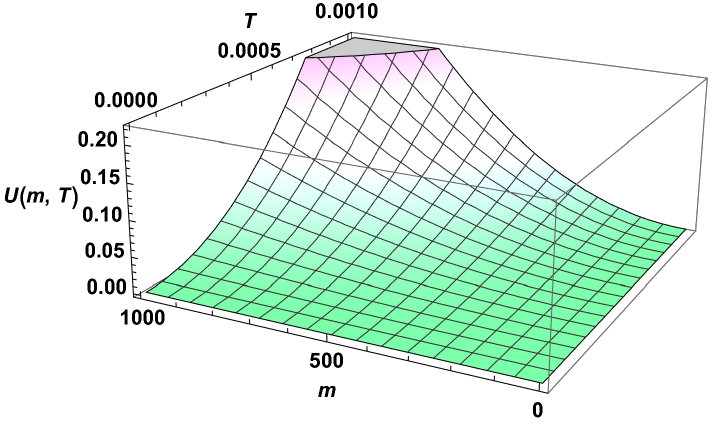}
  \caption{The mean energy for different values of temperature maintaining the same magnitude of mass}\label{ufordifferenttemperature}
\end{figure}

Here, we present the main features ascribed to the mean energy concerning the massive particles. The calculation of this quantity is so important, among other features, because one can correlate, for instance, the work felt by the system with the total amount of heat within it. In other words, with it, we can infer about the well-known first law of thermodynamics. Moreover, another interesting facet to this thermal quantity is the obtainment of the spectral radiance directly from its integrating. To proceed these calculations, we write 
\ie
\begin{split}
U(l_{P},m,\beta) = & \int_{0}^{\infty} \frac{E}{64} \left(\frac{\left(2-4 E l_P\right) \left(2 E-2 E^2 l_P\right)}{\sqrt{\left(2 E-2 E^2 l_P\right){}^2+8 m^2}}-4 E l_P+2\right) \\
& \left(2 E+\sqrt{\left(2 E-2 E^2 l_P\right){}^2+8 m^2}-2 E^2 l_P\right)^{2} \frac{e^{-\beta  E}}{\left(1-e^{-\beta  E}\right)} \,\mathrm{d}E.
\end{split}
\fe
Analogously to what we have carried out to the equation of states, the limit $\left(2 E-2 E^2 l_P\right){}^2 \ll 1$ is also considered. After this consideration, the mean energy is calculated in an \textit{analytical} way:
\ie
\begin{split}
U(l_{P},m,\beta) = & \frac{1}{10080 \beta ^7 \sqrt{m^2}} \left(-4 \sqrt{m^2}-\sqrt{2}\right) \left[126 \beta ^2 l_P \left(\pi ^4 \sqrt{2} \beta  \sqrt{m^2}+20 \beta ^2 m^2 \zeta (3)+240 \zeta (5)\right) \right.\\
& \left. -40 \beta  l_P^2 \left(756 \sqrt{2} \beta  \sqrt{m^2} \zeta (5)+5 \pi ^6\right)+453600 \zeta (7) l_P^3 \right. \\
& \left. -21 \beta ^3 \left(5 \pi ^2 \beta ^2 m^2+60 \sqrt{2} \beta  \sqrt{m^2} \zeta (3)+\pi ^4\right)\right].
\end{split}
\fe
We devote our attention to perform the verification of how mass changes the mean energy; such analysis is displayed in Fig. \ref{ufordifferentmass}. From it, we realize that there is no significant modifications in the shape of the plot to diverse values of mass keeping the same temperature. Furthermore, the behavior of the system for distinct magnitudes of temperatures is another topic worth exploring. In contrast, when we consider the change of the values of temperature maintaining the magnitude of mass constant, we have Fig. \ref{ufordifferenttemperature}. In this configuration of the system, we clearly see that if the range of temperature is $T \leq 1$ GeV, the system indicates instability.

\subsection{The spectral radiance}

\begin{figure}[tbh]
  \centering
  \includegraphics[width=8cm,height=5cm]{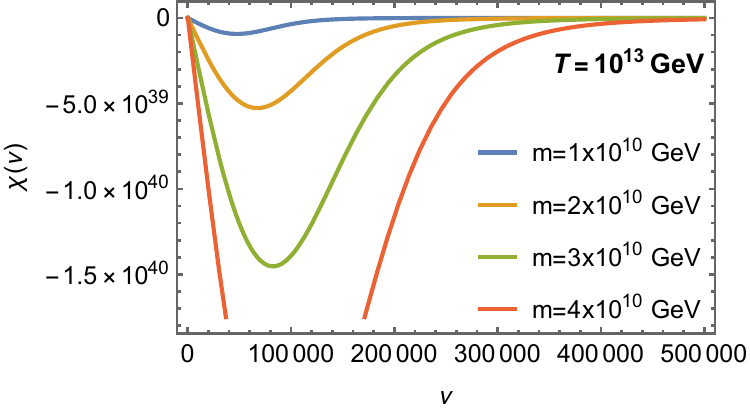}
  \includegraphics[width=8cm,height=5cm]{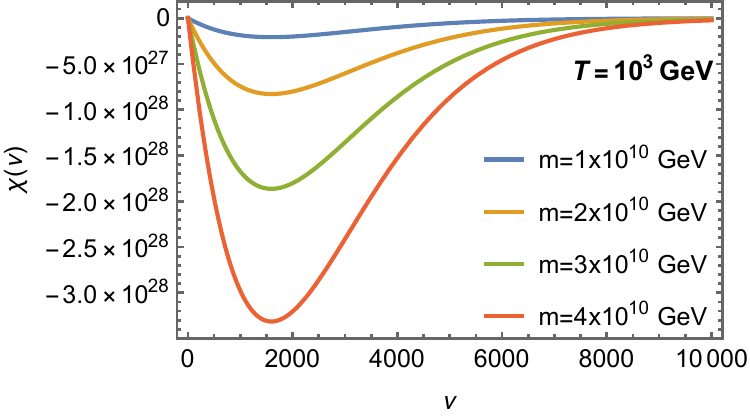}
  \includegraphics[width=8cm,height=5cm]{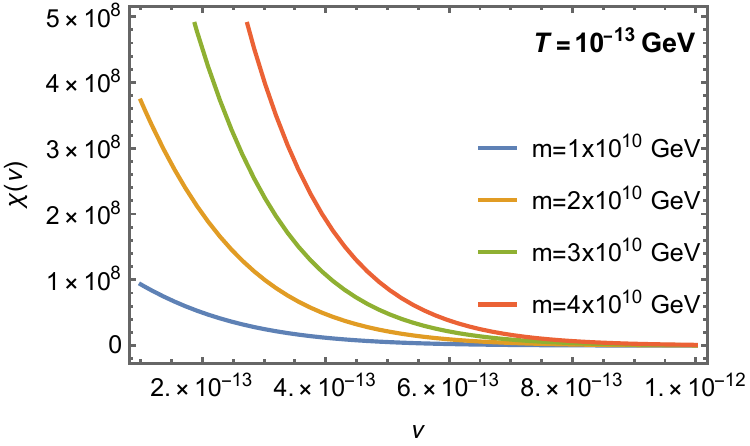}
  \caption{The spectral radiance for the massive case}\label{spectralradiance}
\end{figure}

The spectral radiance is given by
\ie
\begin{split}
\chi(m,\beta,\nu) = &\frac{h \nu}{64} \left(\frac{\left(2-4 h \nu l_P\right) \left(2 h \nu-2 (h \nu)^{2} l_P\right)}{\sqrt{\left(2 h \nu-2 (h \nu)^2 l_P\right){}^2+8 m^2}}-4 h \nu l_P+2\right) \\
& \left(2 h \nu + \sqrt{\left(2 h \nu-2 (h \nu)^{2} l_P\right){}^2+8 m^2}-2 (h \nu)^2 l_P\right)^{2} \frac{e^{-\beta  h \nu}}{\left(1-e^{-\beta  h \nu}\right)},
\end{split}
\fe
where we have considered $E= h \nu$, with $\nu$ being the frequency. The behavior of the well-known spectral radiance is exhibited in Fig. \ref{spectralradiance} for three different regimes of temperature of the universe, e.g., $T= 10^{13}$ GeV (inflationary era), $T= 10^{3}$ GeV (electroweak epoch), and $T= 10^{-13}$ GeV (cosmic microwave background). Note that the first two sets of temperature show a perfect shape of the black body radiation up to a sing; nevertheless, to these cases, the system also indicates instability, as one could naturally expect from the results derived in the last subsection. On the other hand, the last configuration indicates that the shape of the plot is close to that one shown in the Wien’s energy density distribution instead. The next thermodynamic function worth studying is the entropy. To provide such an analysis, we devote the next subsection to it.

\subsection{Entropy}

\begin{figure}[tbh]
  \centering
  \includegraphics[width=8cm,height=5cm]{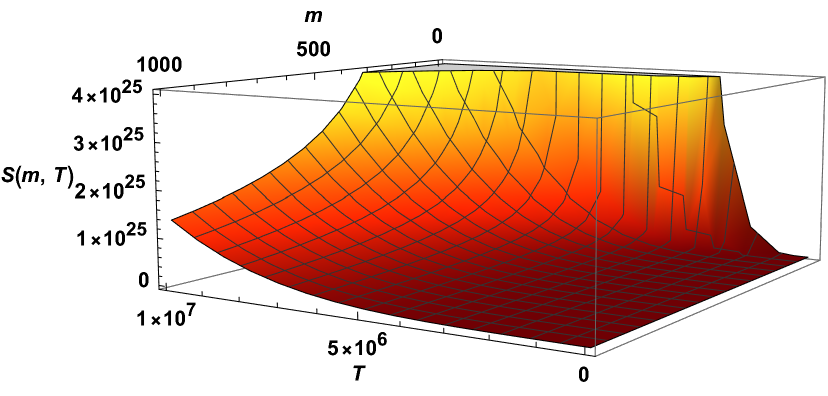}
  \includegraphics[width=8cm,height=5cm]{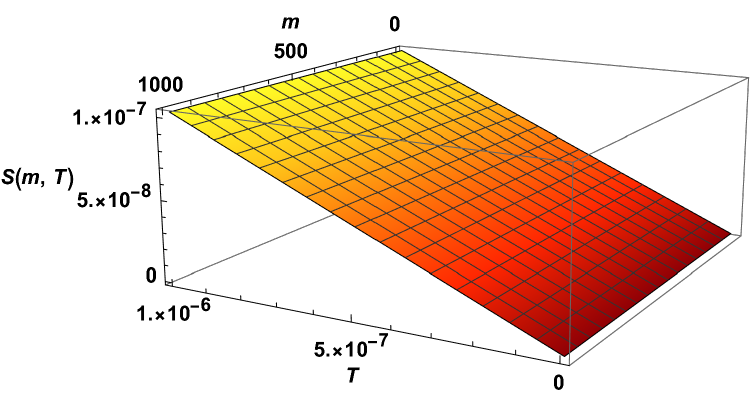}
  \includegraphics[width=8cm,height=5cm]{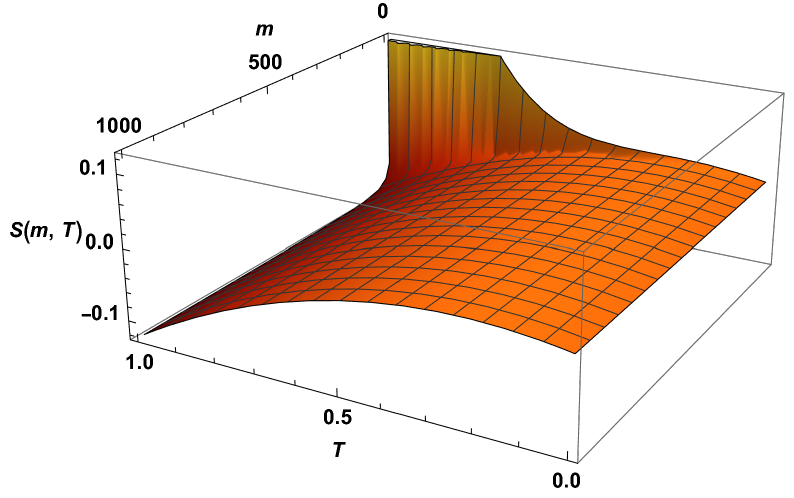}
  \includegraphics[width=8cm,height=5cm]{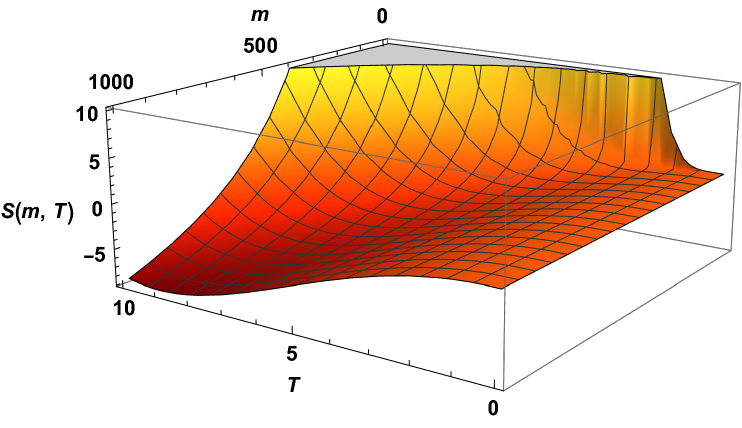}
  \caption{The entropy for the massive case when the mass is kept the same}\label{entropymassivesamemass}
\end{figure}

\begin{figure}[tbh]
  \centering
  \includegraphics[width=8cm,height=5cm]{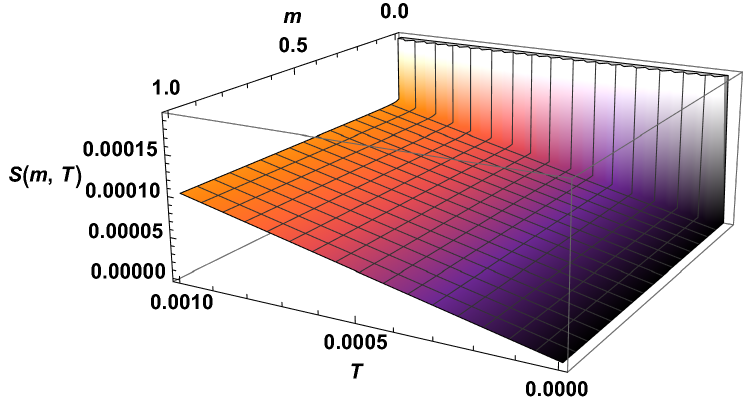}
  \includegraphics[width=8cm,height=5cm]{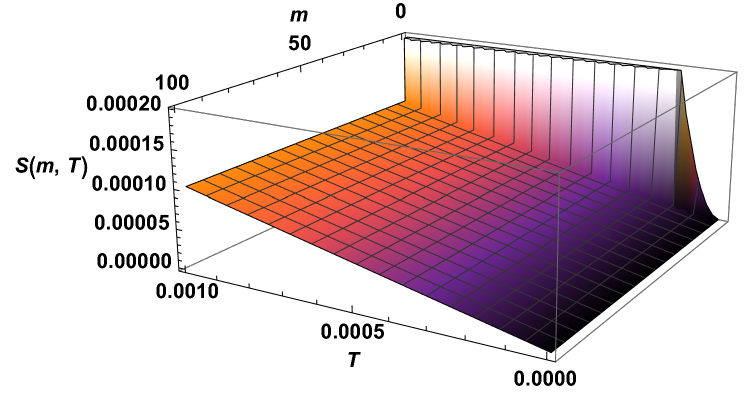}
  \includegraphics[width=8cm,height=5cm]{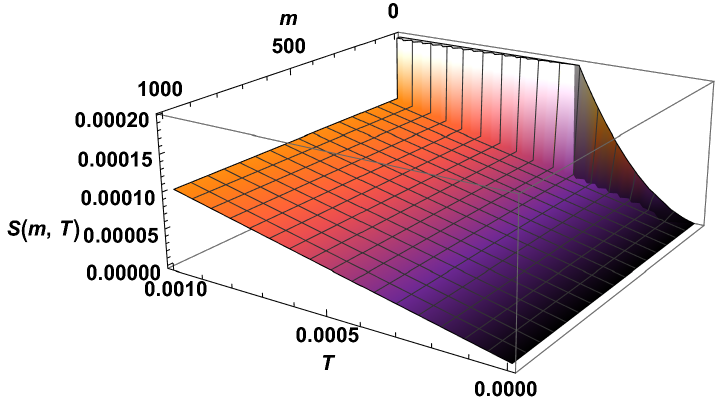}
  \includegraphics[width=8cm,height=5cm]{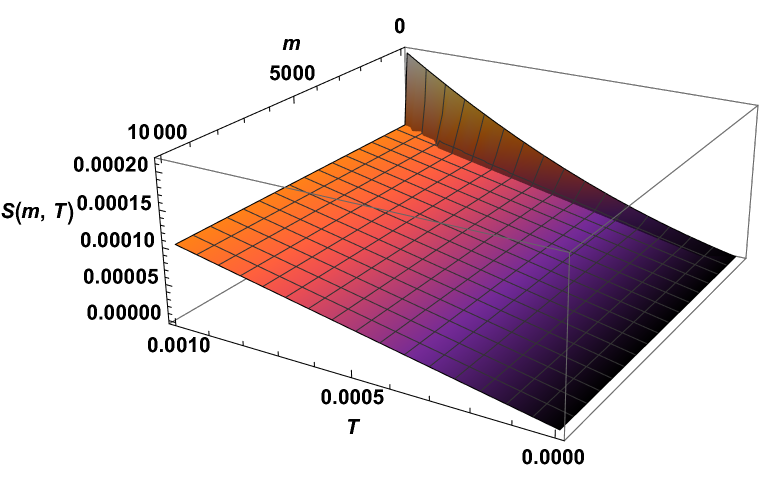}
  \caption{The entropy for the massive case when the temperature is kept the same}\label{entropymassivesamemass2}
\end{figure}

The necessity of having information about the entropy encountered in the system precedes itself. The knowledge of such thermal quantity is notable since we are certainly able to address the behavior of the constituents of matter in a proper manner-- the second law of thermodynamics. In this sense, the entropy is written as
\ie
\begin{split}
S(l_{P},m,\beta) = \int_{0}^{\infty} -\frac{1}{\beta^{2}} & \left[ \frac{\ln \left(1-e^{-\beta  E}\right) \left(\frac{\left(2-4 E l_P\right) \left(2 E-2 E^2 l_P\right)}{\sqrt{\left(2 E-2 E^2 l_P\right){}^2+8 m^2}}-4 E l_P+2\right) }{64 \beta ^2} \right. \\
& \left. \times \frac{\left(2 E+\sqrt{\left(2 E-2 E^2 l_P\right){}^2+8 m^2}-2 E^2 l_P\right){}^2}{64 \beta ^2} \right. \\
& \left. -\frac{E e^{-\beta  E} \left(\frac{\left(2-4 E l_P\right) \left(2 E-2 E^2 l_P\right)}{\sqrt{\left(2 E-2 E^2 l_P\right){}^2+8 m^2}}-4 E l_P+2\right) }{64 \beta  \left(1-e^{-\beta  E}\right)} \right. \\
& \left. \times \frac{\left(2 E+\sqrt{\left(2 E-2 E^2 l_P\right){}^2+8 m^2}-2 E^2 l_P\right){}^2}{64 \beta  \left(1-e^{-\beta  E}\right)} \right] \mathrm{d}E.
\end{split}
\fe
It is important to mention that above expression does not possess an \textit{analytical} solution. In order to carry out this integral in an exact way, we use the limit where  $\left(2 E-2 E^2 l_P\right){}^2 \ll 1$. With this assumption, we get
\ie
S(l_{P},m,\beta) = \frac{-4 \beta  l_P \left(45 \beta  \sqrt{m^2} \zeta (3)+\pi ^4 \sqrt{2}\right)+900 \sqrt{2} \zeta (5) l_P^2+5 \beta ^2 \left(2 \pi ^2 \beta  \sqrt{m^2}+9 \sqrt{2} \zeta (3)\right)}{960 \beta ^4 \sqrt{m^2}}.
\fe
With the purpose of verifying how this thermodynamic function behaves in terms of $m$, and $T$, we provide a plot which is displayed in Fig. \ref{entropymassivesamemass}. Here, the mass is kept constant while we vary the temperature. The system turns out to be sensitive to the change of the values of temperature in a prominent manner. More so, it is worthy to be noted that when the temperature lies between the range of $0 \leq T \leq 1$ -- whenever $m = 10^{3}$ GeV --, the system seems to possess instability. On the other hand, another interesting aspect naturally arises: what would be the comportment of the entropy if different values of mass were taken into account? To answer this question, we properly provide Fig. \ref{entropymassivesamemass2}. Again, the system is quite sensitive to the modification of mass. Note that it tends to have a shape in the form of a sheet for huge values of $m$.

\subsection{The heat capacity}

\begin{figure}[tbh]
  \centering
  \includegraphics[width=8cm,height=5cm]{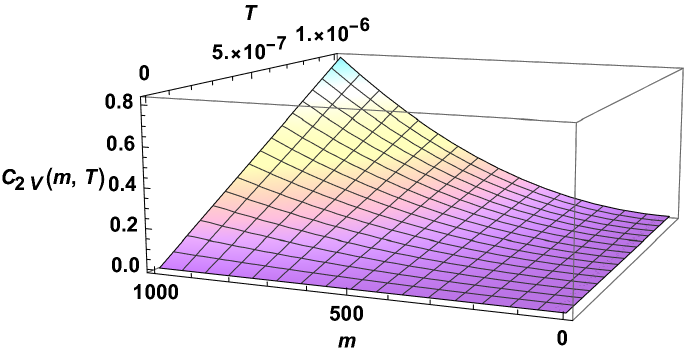}
  \includegraphics[width=8cm,height=5cm]{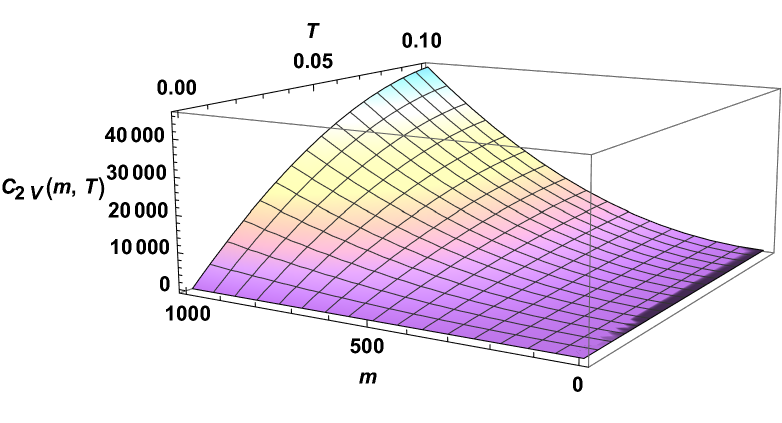}
  \includegraphics[width=8cm,height=5cm]{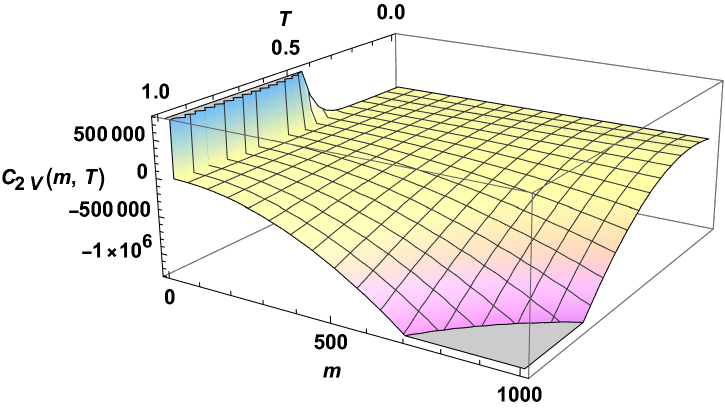}
  \includegraphics[width=8cm,height=5cm]{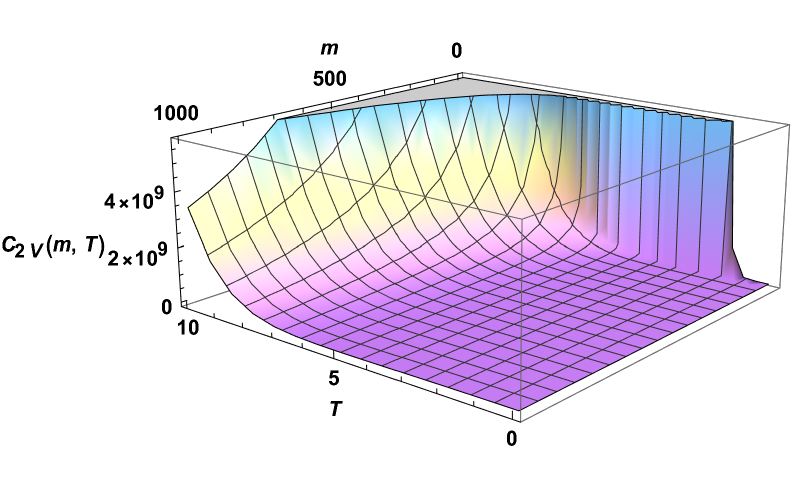}
  \caption{The heat capacity for different values of temperature while the mass is maintained constant}\label{heatcapacitymassivemcte}
\end{figure}

\begin{figure}[tbh]
  \centering
  \includegraphics[width=8cm,height=5cm]{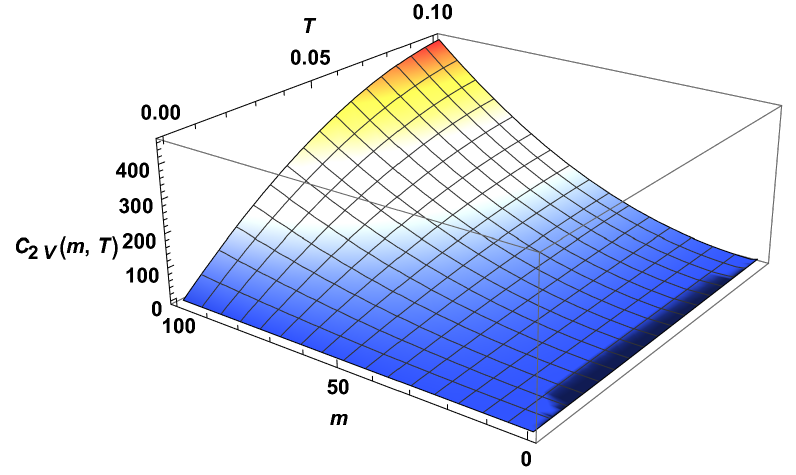}
  \includegraphics[width=8cm,height=5cm]{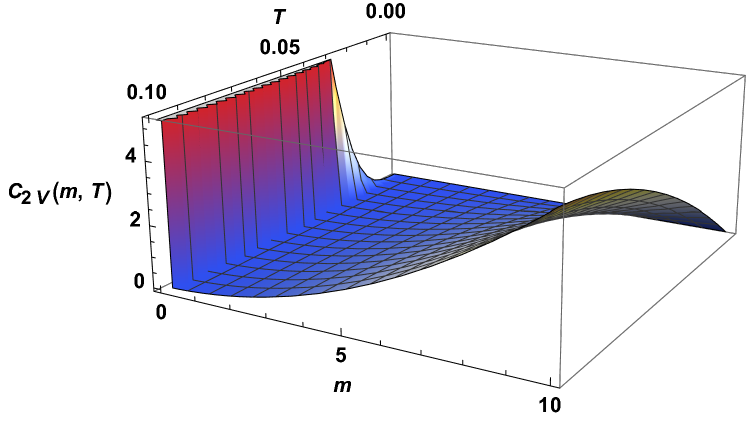}
  \includegraphics[width=8cm,height=5cm]{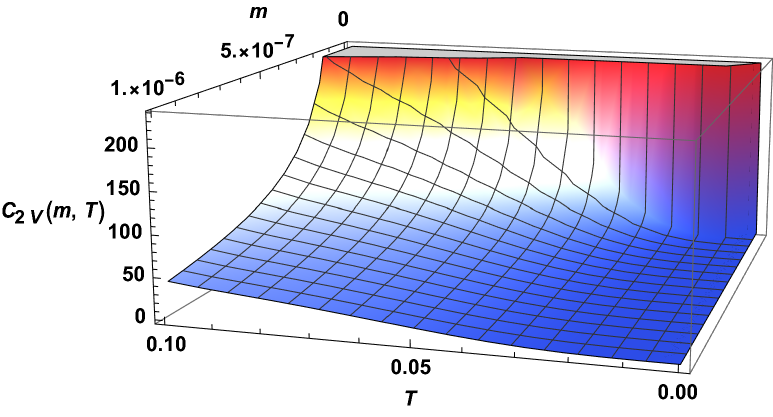}
  \includegraphics[width=8cm,height=5cm]{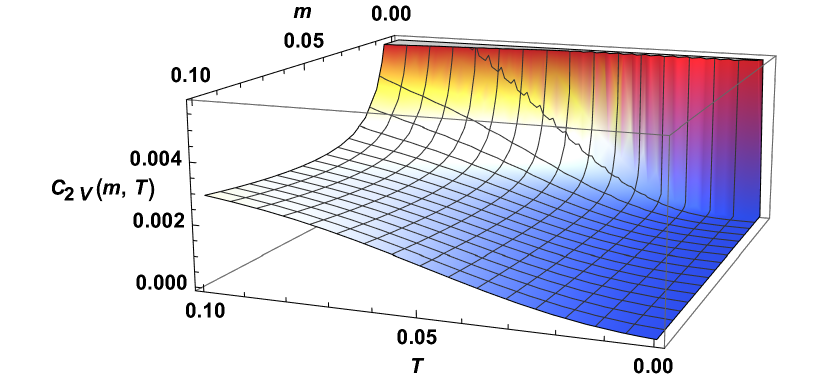}
  \caption{The heat capacity for different values of mass while the temperature is maintained constant}\label{heatcapacitymassivetcte}
\end{figure}

Being an inherent property of any arbitrary substance, the heat capacity is, summarily, the amount of heat energy required to raise its temperature by one unit. Thereby, it is worth knowing such aspect in the context of massive particles in order to have a better comprehension of how mass affects such a system. Then, the heat capacity is given by
\ie
\begin{split}
C_{V} = \int^{\infty}_{0} \beta^{2} & \left[ \frac{E^2 e^{-\beta  E} \left(\frac{\left(2-4 E l_P\right) \left(2 E-2 E^2 l_P\right)}{\sqrt{\left(2 E-2 E^2 l_P\right){}^2+8 m^2}}-4 E l_P+2\right)}{64 \left(1-e^{-\beta  E}\right)} \right. \\
& \left. \times -\frac{\left(2 E+\sqrt{\left(2 E-2 E^2 l_P\right){}^2+8 m^2}-2 E^2 l_P\right){}^2}{64 \left(1-e^{-\beta  E}\right)} \right. \\ 
& \left. \frac{E^2 e^{-2 \beta  E} \left(\frac{\left(2-4 E l_P\right) \left(2 E-2 E^2 l_P\right)}{\sqrt{\left(2 E-2 E^2 l_P\right){}^2+8 m^2}}-4 E l_P+2\right)}{64 \left(1-e^{-\beta  E}\right)^2} \right. \\
&  \left.\times -\frac{\left(2 E+\sqrt{\left(2 E-2 E^2 l_P\right){}^2+8 m^2}-2 E^2 l_P\right){}^2}{64 \left(1-e^{-\beta  E}\right)^2}   \right] \mathrm{d}E.
\end{split}
\fe
Here, the same fact occurred to this thermal function: there is no \textit{analytical} solution. However, after applying the limit $\left(2 E-2 E^2 l_P\right){}^2 \ll 1$, we obtain 
\ie
\begin{split}
C_{V} = &  \frac{1}{420 \beta ^8 \sqrt{m^2}} \left[ 300 \beta ^2 l_P^2 \left(2 \pi ^6 \beta  \sqrt{m^2}+63 \sqrt{2} \left(2 \beta ^2 m^2 \zeta (5)+63 \zeta (7)\right)\right) \right.\\ 
& \left. -2 \beta ^3 l_P \left(63 \pi ^4 \sqrt{2} \beta ^2 m^2+630 \beta  \sqrt{m^2} \left(\beta ^2 m^2 \zeta (3)+60 \zeta (5)\right)+50 \pi ^6 \sqrt{2}\right) \right. \\
& \left. -784 \beta  l_P^3 \left(2025 \beta  \sqrt{m^2} \zeta (7)+\pi ^8 \sqrt{2}\right)+19051200 \sqrt{2} \zeta (9) l_P^4 \right. \\
& \left. + 7 \beta ^4 \left(5 \pi ^2 \beta ^3 \left(m^2\right)^{3/2}+6 \pi ^4 \beta  \sqrt{m^2}+45 \sqrt{2} \left(3 \beta ^2 m^2 \zeta (3)+10 \zeta (5)\right)\right) \right].
\end{split}
\fe
Analogously to what we did with the previous thermodynamic state quantities, we provide an analysis based on fixed values of mass. Such behavior is shown in Fig. \ref{heatcapacitymassivemcte}. The intriguing feature here is that, when the temperature lies in $0<T<6$ GeV, the system brings out instability. On the other hand, when the temperature is kept constant, we have the behavior shown in Fig. \ref{heatcapacitymassivetcte}. In this case, the mass also plays an important role in this thermodynamical property of our system under consideration.

\section{The massless case}

In this section, we focus exclusively on the features ascribed to the massless case. As we did before, the same procedure will be implemented here. In other words, we seek for the accessible state of the system in order to build up the partition function of our ensemble of particles.
In this sense, using the same dispersion relation pointed out previously,
\ie
E \simeq {\bf{k}} + \frac{m^{2}}{2{\bf{k}}} + \alpha l_{P} E^{2},
\nonumber\fe
we may consider the case where $m \rightarrow 0$. Thereby, the dispersion relation related to the massless case is obtained in a straightforward manner:
\ie
E \simeq {\bf{k}} + \alpha l_{P} E^{2}.
\fe
With this, the accessible states can be derived after integrating over the momentum space as follows
\ie
\Omega_{2}(E) = \int_{0}^{\infty} (1-2 E l_{P}) \left(E-E^2 l_{P}\right)^2 \mathrm{d} E.
\fe
In possession of this quantity, the obtainment of the partition function turns out be a straightforward task
\ie
\ln [Z_{2}(\beta)] = -\int_{0}^{\infty} (1-2 E l_{P}) \left(E-E^2 l_{P}\right)^2 \ln \left(1-e^{-\beta  E}\right) \mathrm{d} E.
\fe
As it is well-known from the literature, after knowing the partition function, we are properly able to address the complete behavior of our thermal system and, therefore, all the thermodynamic functions may be obtained. Likewise, in this section, we provide the same thermal quantities calculated to the massive case, i.e., the equation of states, the spectral radiance, the entropy, and the heat capacity. Notably, without any particular limit, the results to this case are carried out in an \textit{analytical} way. It is worth mentioning that \textit{analytical} results involving higher Lorentz-violating scenarios are quite rare even in the context of certain specific approximations. Furthermore, in order to continue our investigation, we initially start with the equation of states.


\subsection{Equation of states}

\begin{figure}[tbh]
  \centering
  \includegraphics[width=9cm,height=6cm]{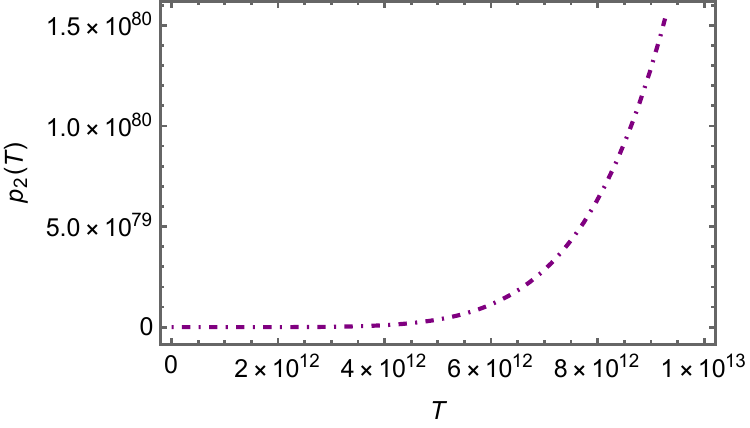}
  \caption{The equation of states for different values of temperature.}\label{equationofstatesmasless}
\end{figure}

When one desires to know about the main aspects of given specific thermodynamical system under consideration, inevitably, one stumbles upon the study of the equation of states. Fundamentally, this occurs because it relates some physical state quantities, such as, pressure, temperature, and volume. Nevertheless, as we shall see, concerning the massless case, the main features is that it correlates the main thermodynamical functions with the temperature and the Riemann zeta function. In this way, we have
\ie
p_{2} = \int_{0}^{\infty} (1-2 E l_{P}) \left(E-E^2 l_{P}\right)^2 \ln \left(1-e^{-\beta  E}\right) \mathrm{d} E,
\fe
which results
\ie
p_{2}(T) = \frac{-7 \pi ^4 T^3 + 75600 l_{P}^3 \zeta (7)-40 \pi ^6 T^{5}  l_{P}^2+7560 T^{4} l_{P} \zeta (5)}{315}.
\fe
Clearly, we see that the pressure has its dependency concerning only on two variables: the temperature $T$, and the Riemann zeta function $\xi(s)$. For the sake of a better comprehension to the reader, we plot the equation of states in Fig. \ref{equationofstatesmasless}.
Note that there exists a huge increment of pressure for different values of temperature. Thereby, the system under consideration, i.e., without mass, up to some specific sets, indicates stability since the results are agreement with the literature-- the second law of the thermodynamics is maintained. Another remarkable information to know about our system is, undeniably, the verification of how the mean energy behaves when mass is no longer invoked. 


\subsection{The mean energy}

In order to provided a better comprehension of the thermodynamical aspects in our investigation, specially, to infer about the amount of work and heat exchanged by the system, it is resealable to have the knowledge of the mean energy. In other words, in possession with it, the first law of the thermodynamics can naturally be addressed. In this sense, we proceed forward writing the proper expression to the mean energy as follows:

\ie
U_{2} = \int_{0}^{\infty} \frac{E e^{-\beta  E} (1-2 E l_{P}) \left(E-E^2 l_{P}\right)^2}{1-e^{-\beta  E}} \mathrm{d}E,
\fe
where, after evaluating above expression, we obtain
\ie
U_{2}(T) = \frac{\pi^4 T^{4}}{15} - 1440 l_{P}^3 \zeta (7)T^{7}+\frac{40 \pi^6 l_{P}^2 T^{6}}{63} - 96 l_{P} \zeta (5) T^{5}.
\fe
\begin{figure}[tbh]
  \centering
  \includegraphics[width=9cm,height=6cm]{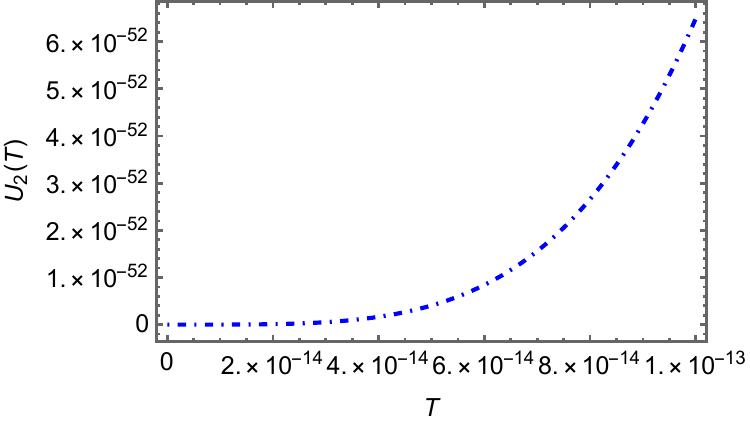}
  \caption{The mean energy for different values of temperature.}\label{meanenergy2massless}
\end{figure}
For the sake of elucidating the comportment of this quantity, we provide the plot shown in Fig. \ref{meanenergy2massless}. Here, we clearly see that the first law and the second law of thermodynamics are preserved considering this simpler case, the massless case. Moreover, there exists another important feature that may be derived from the knowledge of the mean energy, the spectral radiance. Thereby, we examine this physical quantity in the next subsection.


\subsection{The spectral radiance}

\begin{figure}[tbh]
  \centering
  \includegraphics[width=8cm,height=5cm]{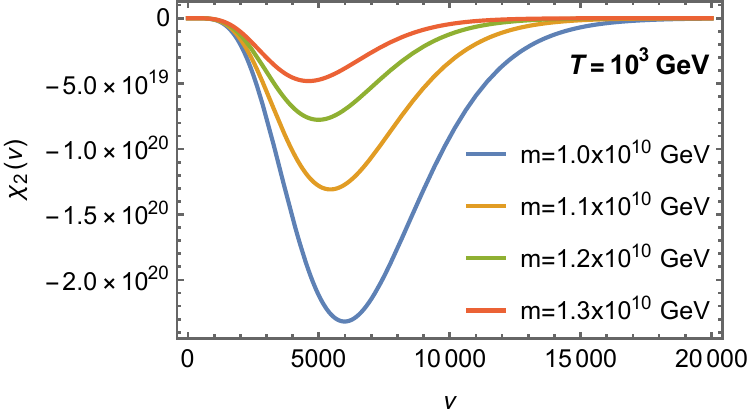}
  \includegraphics[width=8cm,height=5cm]{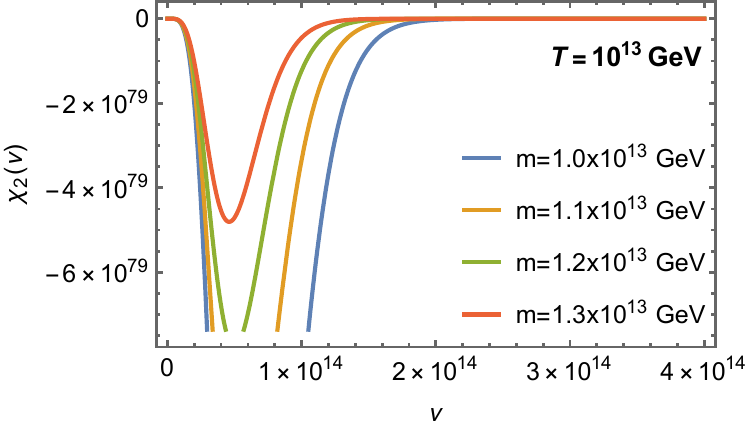}
  \includegraphics[width=8cm,height=5cm]{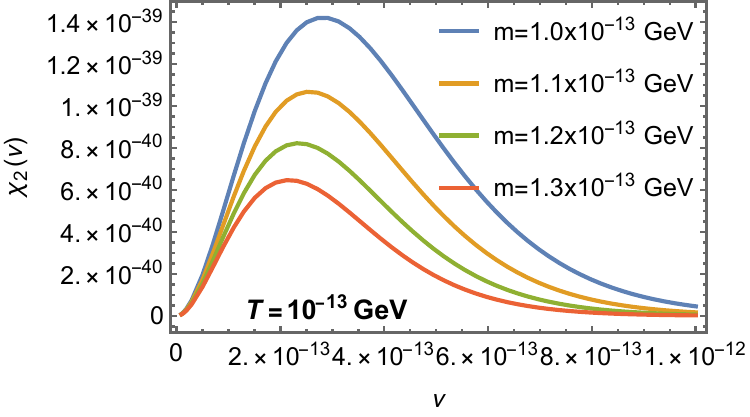}
  \caption{The spectral radiance for the massless case}\label{spectralradiancemassless}
\end{figure}

Here, we supply the study of the black body radiation which is written as 
\ie
\chi_{2}(\beta,\nu) = \frac{ h \nu e^{-\beta  h \nu} (1-2 h \nu l_{P}) \left(h \nu - (h \nu)^2 l_{P}\right)^2}{1-e^{-\beta  h \nu}}.
\fe
Notice that, to this present massless case, the same feature encountered in the massive case remains: the possibility of instabilities when either the electroweak epoch ($T=10^{3}$ GeV) or the inflationary era ($T=10^{13}$ GeV) regime of temperature of the universe are taken into account. Remarkably, the stability gives rise to when the cosmic microwave background regime of temperature ($T= 10^{-13}$ GeV) is considered instead. A detailed representation of these particularities of the spectral radiance is exhibited in Fig. \ref{spectralradiancemassless}.


\subsection{Entropy}

\begin{figure}[tbh]
  \centering
  \includegraphics[width=9cm,height=6cm]{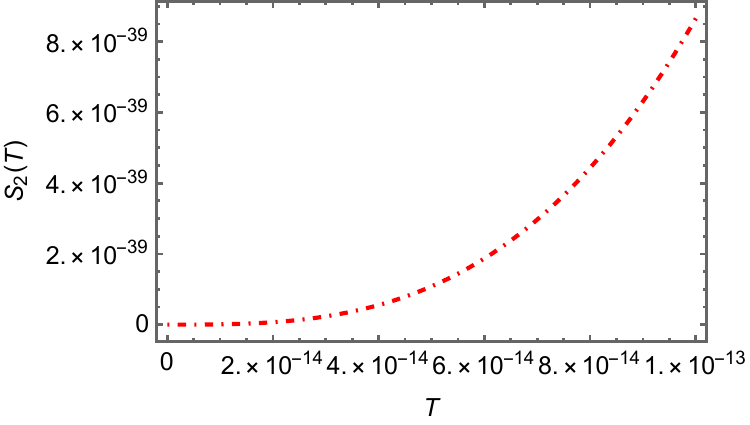}
  \caption{The entropy for different values of temperature.}\label{entropy2massless}
\end{figure}

One of the main concerns of knowing this thermodynamical state quantity is ensuring weather the process of reversibility is present in our system or not. In other words, such information lies in the core of the second law of the thermodynamics. In this way, in order to provide such an analysis, we write the entropy to the massless case as
\ie
S_{2} = \int_{0}^{\infty} -\beta ^2 \left(\frac{(1-2 E l_{P}) \left(E-E^2 l_{P}\right)^2 \ln \left(1-e^{-\beta  E}\right)}{\beta ^2}-\frac{E e^{-\beta  E} (1-2 E l_{P}) \left(E-E^2 l_{P}\right)^2}{\beta  \left(1-e^{-\beta  E}\right)}\right) \mathrm{d}E.
\fe
Notice that this integral can be evaluated without considering any particular limit. Therefore, we have
\ie
S_{2}(T) = \frac{4 \left(7 \pi ^4 T^{3} + 60 \pi ^6 l_{P}^2T^{6} - 9450 T^{4}  l_{P} \zeta (5)\right)}{315} - 1680 l_{P}^3 \zeta (7) T^{6}.
\fe
Its whole behavior as a function of temperature is exhibited in Fig. \ref{entropy2massless}. Again, in this case, there exists an accentuated augmentation in the entropy for different values of temperature.
Finally, to complete our analysis, we have to study the comportment of the heat capacity. To do this, we devote the next subsection for the sake of showing the main characteristics to this remaining quantity.

\subsection{The heat capacity}

\begin{figure}[tbh]
  \centering
  \includegraphics[width=9cm,height=6cm]{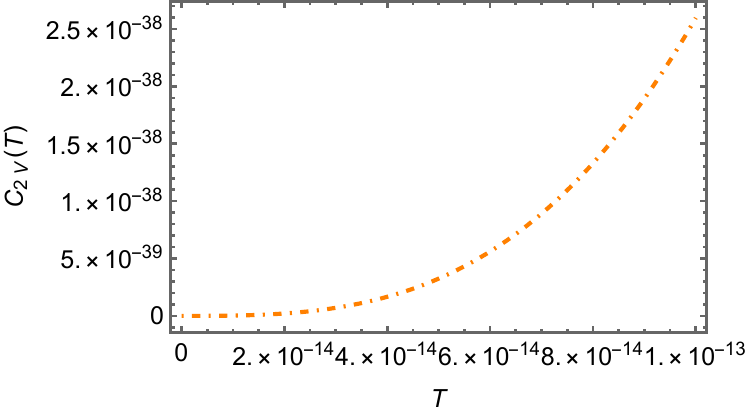}
  \caption{The heat capacity for different values of temperature.}\label{heat2massless}
\end{figure}

For finishing our investigation concerning the thermodynamical properties of the massless case, we present the heat capacity
\ie
C_{2V} = \int_{0}^{\infty} -\beta ^2 \left(-\frac{E^2 e^{-\beta  E} (1-2 E l_{P}) \left(E-E^2 l_{P}\right)^2}{1-e^{-\beta  E}}-\frac{E^2 e^{-2 \beta  E} (1-2 E l_{P}) \left(E-E^2 l_{P}\right)^2}{\left(1-e^{-\beta  E}\right)^2}\right) \mathrm{d}E.
\fe
Analogously to all thermal quantities calculated up to now considering the massless sector, this one has also an \textit{analytical} solution:
\ie
C_{2V}(T) = -\frac{4 \left(-7 \pi ^4 T^3 + 264600 l_{P}^3 \zeta (7)T^{6} - 100 \pi ^6 T^{5}  l_{P}^2+12600 T^{4} l_{P} \zeta (5)\right)}{105}.
\fe
More so, its behavior is shown in Fig. \ref{heat2massless}. Likewise, there is an accentuated increase of its respective curve representing the heat capacity when the temperature increases.

Summarily, as we could verified through the manuscript, the main feature for this simplest massless case is that all the thermal properties follow a similar behavior, i.e., the manner that the plot increases as a function of the temperature $T$ only.


\section{Conclusion}

This work was aimed at analysing the thermodynamical behavior of massive as well as massless particles within the context of Loop Quantum Gravity approach. It was considered a modified dispersion relation which was sufficient to derive all our results in an \textit{analytical} way.
At the beginning, we investigated the massive case where, fundamentally, we verified how the mass modified the thermodynamical functions of the system. On the other hand, in the massless case, the thermodynamical system turned out to depend only on a specific thermal state quantity, the temperature. Furthermore, the examination of the spectral radiation, the equation of states, the mean energy, the entropy and the heat capacity were also accomplished for both cases. Additionally, the thermodynamical properties of our particles under consideration had an explicit dependence on the Riemann zeta function $\xi(s)$.

Within the context of the equation of states, we noticed that, for the massive case, the system was sensitive to the changes of mass. Additionally, in the massless case, we observed that there existed a huge increment of pressure for different values of temperature. Then, the system under consideration, up to some specific sets, indicated stability since the results were in agreement with the literature-- the second law of thermodynamics was maintained.

Concerning the mean energy, we comprehended that there were no substantial modifications in the form of the graphics for different values of mass maintaining a fixed value of temperature. Conversely, when we observed the change of the values of temperature while maintaining constant the magnitude of mass, we clearly perceived that if the range of temperature was $T \leq 1$ GeV, the system would indicate instability. More so, to the massless case, the mean energy had a huge increase in distinct values of temperature.

The spectral radiation to the massive case where performed within the context of three distinct regimes of temperature, i.e., inflationary era ($T = 10^{13}$ GeV), electroweak epoch ($T = 10^{3}$ GeV), and the cosmic microwave background ($T = 10^{-13}$ GeV). In the configuration presented in the first two ones, our results suggested an existence of instability. Nevertheless, only in the latter scenario, the system had a perfect shape of black body radiation. On the other hand, to the massless case, the same regime of temperature was also taken into account. Within the inflationary era ($T = 10^{13}$ GeV) and the electroweak epoch ($T = 10^{3}$ GeV), the system seemed to indicate instability. Irrespective of this, when the cosmic microwave background scheme of temperature ($10^{-13}$ GeV) was invoked, we had the shape of the Wien’s energy density distribution instead.

In the context of the entropy, the system turned out to be responsive to the changes of the values of temperature in a recognizable manner. Furthermore, it was noted if the temperature lied in the range $0 \leq T \leq 1$, i.e., with $m = 10^{3}$ GeV, the system would possess instability. On the other hand, to the massless case, there existed an accentuated augmentation in the entropy behavior for distinct values of temperature.

Finally, to the heat capacity, some intriguing features emerged. For instance, when the temperature lied in $0<T<6$ GeV, the system showed instability. And, considering the massless case instead, the heat capacity increased in a huge way when there was a modification in the temperature.


\section*{Acknowledgments}
\hspace{0.5cm} This work has been supported by Conselho Nacional de Desenvolvimento Cient\'{\i}fico e Tecnol\'{o}gico (CNPq) - 142412/2018-0. Most of the calculations presented in this manuscript were accomplished by using the \textit{Mathematica} software. More so, we would like to thank Izeldin and Mylkleany for the careful reading of this manuscript.

\bibliographystyle{ieeetr}
\bibliography{main}

\begin{thebibliography}{10}

\bibitem{rovelli2004quantum}
C.~Rovelli, {\em Quantum gravity}.
\newblock Cambridge university press, 2004.

\bibitem{kiefer2007quantum}
C.~Kiefer, ``Why quantum gravity?,'' in {\em Approaches to fundamental
  physics}, pp.~123--130, Springer, 2007.

\bibitem{bianchi2011polyhedra}
E.~Bianchi, P.~Dona, and S.~Speziale, ``Polyhedra in loop quantum gravity,''
  {\em Physical Review D}, vol.~83, no.~4, p.~044035, 2011.

\bibitem{carlip2001quantum}
S.~Carlip, ``Quantum gravity: a progress report,'' {\em Reports on progress in
  physics}, vol.~64, no.~8, p.~885, 2001.

\bibitem{amelino2004severe}
G.~Amelino-Camelia, M.~Arzano, and A.~Procaccini, ``Severe constraints on the
  loop-quantum-gravity energy-momentum dispersion relation from the black-hole
  area-entropy law,'' {\em Physical Review D}, vol.~70, no.~10, p.~107501,
  2004.

\bibitem{bekenstein2020black}
J.~D. Bekenstein, ``Black holes and entropy,'' in {\em JACOB BEKENSTEIN: The
  Conservative Revolutionary}, pp.~307--320, World Scientific, 2020.

\bibitem{ghosh2014statistics}
A.~Ghosh, K.~Noui, and A.~Perez, ``Statistics, holography, and black hole
  entropy in loop quantum gravity,'' {\em Physical Review D}, vol.~89, no.~8,
  p.~084069, 2014.

\bibitem{mansuroglu2021fermion}
R.~Mansuroglu and H.~Sahlmann, ``Fermion spins in loop quantum gravity,'' {\em
  Physical Review D}, vol.~103, no.~6, p.~066016, 2021.

\bibitem{xiao2022logarithmic}
Y.~Xiao and Y.~Tian, ``Logarithmic correction to black hole entropy from the
  nonlocality of quantum gravity,'' {\em Physical Review D}, vol.~105, no.~4,
  p.~044013, 2022.

\bibitem{rovelli1996black}
C.~Rovelli, ``Black hole entropy from loop quantum gravity,'' {\em Physical
  Review Letters}, vol.~77, no.~16, p.~3288, 1996.

\bibitem{ashtekar1998quantum}
A.~Ashtekar, J.~Baez, A.~Corichi, and K.~Krasnov, ``Quantum geometry and black
  hole entropy,'' {\em Physical Review Letters}, vol.~80, no.~5, p.~904, 1998.

\bibitem{kaul2000logarithmic}
R.~K. Kaul and P.~Majumdar, ``Logarithmic correction to the bekenstein-hawking
  entropy,'' {\em Physical Review Letters}, vol.~84, no.~23, p.~5255, 2000.

\bibitem{strominger1996microscopic}
A.~Strominger and C.~Vafa, ``Microscopic origin of the bekenstein-hawking
  entropy,'' {\em Physics Letters B}, vol.~379, no.~1-4, pp.~99--104, 1996.

\bibitem{solodukhin1998entropy}
S.~N. Solodukhin, ``Entropy of the schwarzschild black hole and the
  string--black-hole correspondence,'' {\em Physical Review D}, vol.~57, no.~4,
  p.~2410, 1998.

\bibitem{solodukhin2020logarithmic}
S.~N. Solodukhin, ``Logarithmic terms in entropy of schwarzschild black holes
  in higher loops,'' {\em Physics Letters B}, vol.~802, p.~135235, 2020.

\bibitem{meissner2004black}
K.~A. Meissner, ``Black-hole entropy in loop quantum gravity,'' {\em Classical
  and Quantum Gravity}, vol.~21, no.~22, p.~5245, 2004.

\bibitem{kostelecky2011data}
V.~A. Kosteleck{\`y} and N.~Russell, ``Data tables for lorentz and c p t
  violation,'' {\em Reviews of Modern Physics}, vol.~83, no.~1, p.~11, 2011.

\bibitem{colladay2004statistical}
D.~Colladay and P.~McDonald, ``Statistical mechanics and lorentz violation,''
  {\em Physical Review D}, vol.~70, no.~12, p.~125007, 2004.

\bibitem{aa2021lorentz}
A.~A. Ara{\'u}jo~Filho, ``Lorentz-violating scenarios in a thermal reservoir,''
  {\em The European Physical Journal Plus}, vol.~136, no.~4, pp.~1--14, 2021.

\bibitem{araujo2021thermodynamic}
A.~A. Ara{\'u}jo~Filho and R.~V. Maluf, ``Thermodynamic properties in
  higher-derivative electrodynamics,'' {\em Brazilian Journal of Physics},
  vol.~51, no.~3, pp.~820--830, 2021.

\bibitem{anacleto2018lorentz}
M.~Anacleto, F.~Brito, E.~Maciel, A.~Mohammadi, E.~Passos, W.~Santos, and
  J.~Santos, ``Lorentz-violating dimension-five operator contribution to the
  black body radiation,'' {\em Physics Letters B}, vol.~785, pp.~191--196,
  2018.

\bibitem{casana2008lorentz}
R.~Casana, M.~M. Ferreira~Jr, and J.~S. Rodrigues, ``Lorentz-violating
  contributions of the carroll-field-jackiw model to the cmb anisotropy,'' {\em
  Physical Review D}, vol.~78, no.~12, p.~125013, 2008.

\bibitem{casana2009finite}
R.~Casana, M.~M. Ferreira~Jr, J.~S. Rodrigues, and M.~R. Silva, ``Finite
  temperature behavior of the c p t-even and parity-even electrodynamics of the
  standard model extension,'' {\em Physical Review D}, vol.~80, no.~8,
  p.~085026, 2009.

\bibitem{araujo2021higher}
A.~Ara{\'u}jo~Filho and A.~Y. Petrov, ``Higher-derivative lorentz-breaking
  dispersion relations: a thermal description,'' {\em The European Physical
  Journal C}, vol.~81, no.~9, pp.~1--16, 2021.

\bibitem{reis2021thermal}
J.~Reis {\em et~al.}, ``Thermal aspects of interacting quantum gases in
  lorentz-violating scenarios,'' {\em The European Physical Journal Plus},
  vol.~136, no.~3, pp.~1--30, 2021.

\bibitem{petrov2021bouncing}
A.~Y. Petrov {\em et~al.}, ``Bouncing universe in a heat bath,'' {\em arXiv
  preprint arXiv:2105.05116}, 2021.

\bibitem{petrov2021bouncing2}
A.~A. Ara{\'u}jo~Filho and A.~Y. Petrov, ``Bouncing universe in a heat bath,''
  {\em International Journal of Modern Physics A}, vol.~36, no.~34 \& 35,
  (2021) 2150242. DOI: doi.org/10.1142/S0217751X21502420.

\bibitem{aaa2021thermodynamics}
A.~A. Ara{\'u}jo~Filho, ``Thermodynamics of massless particles in curved
  spacetime,'' {\em arXiv preprint arXiv:2201.00066}, 2021.

\bibitem{amelino2001testable}
G.~Amelino-Camelia, ``Testable scenario for relativity with minimum length,''
  {\em Physics Letters B}, vol.~510, no.~1-4, pp.~255--263, 2001.

\bibitem{amelino1998tests}
G.~Amelino-Camelia, J.~Ellis, N.~Mavromatos, D.~V. Nanopoulos, and S.~Sarkar,
  ``Tests of quantum gravity from observations of $\gamma$-ray bursts,'' {\em
  Nature}, vol.~393, no.~6687, pp.~763--765, 1998.

\bibitem{garay1998spacetime}
L.~J. Garay, ``Spacetime foam as a quantum thermal bath,'' {\em Physical Review
  Letters}, vol.~80, no.~12, p.~2508, 1998.

\bibitem{amelino2002doubly}
G.~Amelino-Camelia, ``Doubly-special relativity: first results and key open
  problems,'' {\em International Journal of Modern Physics D}, vol.~11, no.~10,
  pp.~1643--1669, 2002.

\bibitem{magueijo2003generalized}
J.~Magueijo and L.~Smolin, ``Generalized lorentz invariance with an invariant
  energy scale,'' {\em Physical Review D}, vol.~67, no.~4, p.~044017, 2003.

\bibitem{kowalski2003non}
J.~Kowalski-Glikman and S.~Nowak, ``Non-commutative space--time of doubly
  special relativity theories,'' {\em International Journal of Modern Physics
  D}, vol.~12, no.~02, pp.~299--315, 2003.

\bibitem{amelino2002doubly2}
G.~Amelino-Camelia, ``Doubly special relativity,'' {\em arXiv preprint
  gr-qc/0207049}, 2002.

\bibitem{cai2014testing}
Y.-F. Cai and Y.~Wang, ``Testing quantum gravity effects with latest cmb
  observations,'' {\em Physics Letters B}, vol.~735, pp.~108--111, 2014.

\bibitem{myers2004experimental}
R.~C. Myers and M.~Pospelov, ``Experimental challenges for quantum gravity,''
  in {\em Quantum Theory and Symmetries}, pp.~732--744, World Scientific, 2004.

\bibitem{arzano2016gravity}
M.~Arzano and G.~Calcagni, ``What gravity waves are telling about quantum
  spacetime,'' {\em Physical Review D}, vol.~93, no.~12, p.~124065, 2016.

\bibitem{sudarsky2003bounds}
D.~Sudarsky, L.~Urrutia, and H.~Vucetich, ``Bounds on stringy quantum gravity
  from low energy existing data,'' {\em Physical Review D}, vol.~68, no.~2,
  p.~024010, 2003.

\bibitem{calcagni2019gravitational}
G.~Calcagni, S.~Kuroyanagi, S.~Marsat, M.~Sakellariadou, N.~Tamanini, and
  G.~Tasinato, ``Gravitational-wave luminosity distance in quantum gravity,''
  {\em Physics Letters B}, vol.~798, p.~135000, 2019.

\bibitem{chen2014effects}
D.~Chen, H.~Wu, H.~Yang, and S.~Yang, ``Effects of quantum gravity on black
  holes,'' {\em International Journal of Modern Physics A}, vol.~29, no.~26,
  p.~1430054, 2014.

\bibitem{hossenfelder2013minimal}
S.~Hossenfelder, ``Minimal length scale scenarios for quantum gravity,'' {\em
  Living Reviews in Relativity}, vol.~16, no.~1, pp.~1--90, 2013.

\bibitem{mercati2010probing}
F.~Mercati, D.~Maz{\'o}n, G.~Amelino-Camelia, J.~M. Carmona, J.~L. Cort{\'e}s,
  J.~Indur{\'a}in, C.~L{\"a}mmerzahl, and G.~M. Tino, ``Probing the
  quantum-gravity realm with slow atoms,'' {\em Classical and Quantum Gravity},
  vol.~27, no.~21, p.~215003, 2010.

\bibitem{laanemets2022observables}
D.~L{\"a}{\"a}nemets, M.~Hohmann, and C.~Pfeifer, ``Observables from
  spherically symmetric modified dispersion relations,'' {\em arXiv preprint
  arXiv:2201.04694}, 2022.

\bibitem{gong2022gravitational}
C.~Gong, T.~Zhu, R.~Niu, Q.~Wu, J.-L. Cui, X.~Zhang, W.~Zhao, and A.~Wang,
  ``Gravitational wave constraints on lorentz and parity violations in gravity:
  high-order spatial derivative cases,'' {\em Physical Review D}, vol.~105,
  no.~4, p.~044034, 2022.

\bibitem{amelino12004quantum}
G.~Amelino-Camelia, L.~Smolin, and A.~Starodubtsev, ``Quantum symmetry, the
  cosmological constant and planck-scale phenomenology,'' {\em Classical and
  Quantum Gravity}, vol.~21, no.~13, p.~3095, 2004.

\bibitem{alfaro2000quantum}
J.~Alfaro, H.~A. Morales-Tecotl, and L.~F. Urrutia, ``Quantum gravity
  corrections to neutrino propagation,'' {\em Physical Review Letters},
  vol.~84, no.~11, p.~2318, 2000.

\bibitem{smolin2002quantum}
L.~Smolin, ``Quantum gravity with a positive cosmological constant,'' {\em
  arXiv preprint hep-th/0209079}, 2002.

\bibitem{gambini1999nonstandard}
R.~Gambini and J.~Pullin, ``Nonstandard optics from quantum space-time,'' {\em
  Physical Review D}, vol.~59, no.~12, p.~124021, 1999.

\bibitem{ronco2016uv}
M.~Ronco, ``On the uv dimensions of loop quantum gravity,'' {\em Advances in
  High Energy Physics}, vol.~2016, 2016.

\bibitem{bojowald2005loop}
M.~Bojowald, H.~A. Morales-T{\'e}cotl, and H.~Sahlmann, ``Loop quantum gravity
  phenomenology and the issue of lorentz invariance,'' {\em Physical Review D},
  vol.~71, no.~8, p.~084012, 2005.

\bibitem{brahma2017linking}
S.~Brahma, M.~Ronco, G.~Amelino-Camelia, and A.~Marciano, ``Linking loop
  quantum gravity quantization ambiguities with phenomenology,'' {\em Physical
  Review D}, vol.~95, no.~4, p.~044005, 2017.

\bibitem{ashtekar2021short}
A.~Ashtekar and E.~Bianchi, ``A short review of loop quantum gravity,'' {\em
  Reports on Progress in Physics}, 2021.

\bibitem{maluf2019antisymmetric}
R.~Maluf, A.~Ara{\'u}jo~Filho, W.~Cruz, and C.~Almeida, ``Antisymmetric tensor
  propagator with spontaneous lorentz violation,'' {\em EPL (Europhysics
  Letters)}, vol.~124, no.~6, p.~61001, 2019.

\bibitem{schreck2022lorentz}
M.~Schreck, ``Lorentz violation in astroparticles and gravitational waves,''
  2022.

\bibitem{greiner2012thermodynamics}
W.~Greiner, L.~Neise, and H.~St{\"o}cker, {\em Thermodynamics and statistical
  mechanics}.
\newblock Springer Science \& Business Media, 2012.

\bibitem{reis2020does}
A.~Ara{\'u}jo~Filho and J.~Reis, ``How does geometry affect quantum gases?,''
  {\em International Journal of Modern Physics A}, vol.~37, no.~11n12,
  p.~2250071, 2022.

\bibitem{oliveira2019thermodynamic}
R.~R. Oliveira, A.~A. Ara{\'u}jo~Filho, F.~C. Lima, R.~V. Maluf, and C.~A.
  Almeida, ``Thermodynamic properties of an aharonov-bohm quantum ring,'' {\em
  The European Physical Journal Plus}, vol.~134, no.~10, p.~495, 2019.

\bibitem{oliveira2020relativistic}
R.~Oliveira, A.~Ara{\'u}jo~Filho, R.~Maluf, and C.~Almeida, ``The relativistic
  aharonov--bohm--coulomb system with position-dependent mass,'' {\em Journal
  of Physics A: Mathematical and Theoretical}, vol.~53, no.~4, p.~045304, 2020.

\bibitem{oliveira2020thermodynamic}
R.~Oliveira {\em et~al.}, ``Thermodynamic properties of neutral dirac particles
  in the presence of an electromagnetic field,'' {\em The European Physical
  Journal Plus}, vol.~135, no.~1, pp.~1--10, 2020.

\bibitem{reis2021fermions}
A.~Ara{\'u}jo~Filho, J.~Reis, and S.~Ghosh, ``Fermions on a torus knot,'' {\em
  The European Physical Journal Plus}, vol.~137, no.~5, pp.~1--15, 2022.

\end{thebibliography}

\end{document}